# The role of spin-orbit coupling in the electronic structure of IrO$_2$

Pranab Kumar Das,[1,2,▲] Jagoda Sławińska,[3] Ivana Vobornik,[1] Jun Fujii,[1] Anna Regoutz,[4] Juhan M. Kahk,[4] David O. Scanlon,[5,6] Benjamin J. Morgan,[7] Cormac McGuinness,[8] Evgeny Plekhanov,[3,†] Domenico Di Sante,[9] Ying-Sheng Huang,[10,‡] Ruei-San Chen,[11] Giorgio Rossi,[1,12] Silvia Picozzi,[3,*] William R. Branford,[13,] Giancarlo Panaccione[1,*] and David J. Payne[4,*]

1. *Istituto Officina dei Materiali (IOM)-CNR, Laboratorio TASC, in Area Science Park, S.S.14, Km 163.5, I-34149 Trieste, Italy.*
2. *International Centre for Theoretical Physics, Strada Costiera 11, 34100 Trieste, Italy.*
3. *Consiglio Nazionale delle Ricerche (CNR-SPIN), Sede temporánea di Chieti, 66100 Chieti, Italy*
4. *Department of Materials, Imperial College London, Exhibition Road, London SW7 2AZ, United Kingdom.*
5. *University College London, Kathleen Lonsdale Materials Chemistry, Department of Chemistry, 20 Gordon Street, London WC1H 0AJ, United Kingdom.*
6. *Diamond Light Source Ltd., Diamond House, Harwell Science and Innovation Campus, Didcot, Oxfordshire OX11 0DE, United Kingdom.*
7. *Department of Chemistry, University of Bath, Claverton Down Road, Bath BA2 7AY, United Kingdom.*
8. *School of Physics, Trinity College Dublin, College Green, Dublin 2, Ireland.*
9. *Institut für Theoretische Physik und Astrophysik, Universität Würzburg, Am Hubland Campus Süd, Würzburg 97074, Germany.*
10. *Department of Electric Engineering, National Taiwan University of Science and Technology, Taipei 106, Taiwan*
11. *Graduate Institute of Applied Science and Technology, National Taiwan University of Science and Technology, Taipei 106, Taiwan*
12. *Dipartimento di Fisica, Università di Milano, Via Celoria 16, I-20133 Milano, Italy.*
13. *Blackett Laboratory, Department of Physics, Imperial College, Prince Consort Road, South Kensington, London SW7 2AZ, United Kingdom.*

* Corresponding authors: d.payne@imperial.ac.uk; silvia.picozzi@spin.cnr.it giancarlo.panaccione@elettra.eu
Current addresses: ▲Singapore Synchrotron Light Source, National University of Singapore, 5 Research Link, Singapore - 117603. † King's College London, Theory and Simulation of Condensed Matter (TSCM), The Strand, London WC2R 2LS, United Kingdom. ‡ Deceased

**Abstract**

The delicate interplay of electronic charge, spin, and orbital degrees of freedom is in the heart of many novel phenomena across the transition metal oxide family. Here, by combining high-resolution angle resolved photoemission spectroscopy and first principles calculations (with and without spin-orbit coupling), the electronic structure of the rutile binary iridate, $IrO_2$ is investigated. The detailed study of electronic bands measured on a high-quality single crystalline sample, and use of a wide range of photon energy provide a huge improvement over the previous studies. The excellent agreement between theory and experimental results shows that the single-particle DFT description of $IrO_2$ band structure is adequate, without the need of invoking any treatment of correlation effects. Although many observed features point to a 3D nature of the electronic structure, clear surface effects are revealed. The discussion of the orbital character of the relevant bands crossing the Fermi level sheds light on spin orbit coupling-driven phenomena in this material, unveiling a spin-orbit induced avoided crossing, a property likely to play key role in its large spin Hall effect.

**Introduction**

Understanding the influence of spin-orbit coupling (SOC) in 5$d$ transition metal oxides (TMOs) is a dynamic and challenging area of current condensed matter physics research [1-5], as the interplay of heavy elements, peculiar symmetry properties and exotically textured band structures are key ingredients in determining several novel phenomena. Transition metal oxides are known to reveal a wide range of exotic states including spin-orbit-assisted Mott insulator [3, 6], spin Hall effect [7], spin liquid behavior [8-11], topological Weyl semimetallic state [12, 13], all primarily driven by SOC blended together with unique nature of $d$ shell electrons [14]. The competing balance between the localized electronic behavior within the TM-O$_6$ octahedral building blocks and the delocalized behavior among these units when they form a crystal matrix, makes TMOs so diverse and fascinating. Among the large family of 5d TMOs, the iridates represent a particularly interesting class. Their research so far has been mainly focused on the role of SOC related to the metal-insulator transition [6, 8, 15], unusual magnetic ground states [9, 16] or the possible existence of topological conductivity [12, 17, 18]. Intriguingly, even a parent compound representing the simplest binary oxide of iridium, IrO$_2$, exhibits several unusual properties. K. Fujiwara *et al* [7] have recently reported a very large spin Hall resistivity in this material which makes it a promising candidate as an efficient spin detector. A recent theoretical study by Sun *et al.* [19], has revealed IrO$_2$ as a topological nodal line semimetal, with two distinct types of nodal lines and that the excellent spin detection properties arise from the existence of these topological nodal lines in the 3D Brillouin zone (BZ). They postulate that the first (second) type of Dirac nodal line (DNL) is characterized by a non-trivial finite (trivial zero) Berry phase calculated along a Wilson loop encircling the line. Further spin Berry curvature calculations confirmed that it is the first type of DNL that strongly contributes to the spin Hall conductivity. In the light of this, the experimental corroboration of the spin-orbit derived features in the electronic structure seems to be extremely important for the future development of this field. Here, we aim to identify signatures of SOC in the electronic bands, and discuss their possible role behind the large spin Hall effect of IrO$_2$.

We investigated the electronic structure of IrO$_2$ by combining the high-resolution angle-resolved photoelectron spectroscopy (ARPES) with density functional theory (DFT) and Green's functions calculations, which allowed us to identify features strongly dependent on the $k$ vector normal to the (110) surface, i.e. a 3D character, coexisting with both undispersed bands and prominent surface effects/states. Moreover, photon energy dependent analysis of the orbital character of states near the Fermi level allows to reveal the presence of an avoided crossing at the high-symmetry hextuple point along the $\Gamma - Z$ direction in the Brillouin zone [19]; we discuss this observation as possibly connected to the spin Hall properties of IrO$_2$.

**Methods**

The black metallic oxide $IrO_2$ adopts the rutile structure with the space group *P4$_2$/mnm* and lattice parameters *a* = *b* = 4.5050 Å and *c* = 3.1586 Å [20, 21]. The $IrO_2$ crystals were grown by a chemical vapor transport technique in a flow oxygen system with oxygen pressure greater than one atmosphere [22]. Further details can be found in Ref. [23]. The single crystal of $IrO_2$ was cleaved mechanically inside UHV to obtain the (110) surface. High resolution ARPES measurements are performed using ScientaOmicron DA30 electron analyzer at 78 K (liquid nitrogen cooled cryostat exploiting linearly polarized synchrotron radiation of 40 – 85 eV at an angular and energy resolution better than 0.2° and 20 meV respectively. The experiments were performed at the APE-IOM beamline at the ELETTRA Sincrotrone, Trieste [24]. The sample surface was exposed to a pressure lower than 1 x $10^{-10}$ mbar during the ARPES measurements.

DFT calculations were performed by using the GREEN code [25] interfaced with the SIESTA package [26]. The exchange and correlation terms were considered within the GGA in the Perdew–Burke–Ernzerhof formalism [27]. Core electrons were replaced by norm-conserving pseudopotentials of the Troulliers-Martin type, with core corrections included for Ir atoms. The atomic orbital (AO) basis set consisted of double-zeta polarized numerical orbitals strictly localized. The confinement energy in the basis generation process was set to 100 meV. Real space three-center integrals were computed over 3D-grids with a resolution equivalent to 500 Rydbergs mesh cut-off. The temperature $K_BT$ in the Fermi–Dirac distribution was set to 10 meV. Spin–orbit coupling has been self-consistently taken into account as implemented in Ref. [28]. To model the surface electronic structure, we considered a slab containing 4 units of $IrO_2$ stacked onto each other along the [110] direction. The electronic structures have been calculated in the form of **k**- and E-resolved projected density of states *p*DOS (**k**, E) following the Green's functions matching technique [29].

**Results**

Figure 1 summarizes the band structures and Fermi surfaces measured by ARPES. The (primitive) bulk and surface Brillouin zones, as well as the convention used to identify the high symmetry lines, are shown in Figures 1(a) and (b). We defined $\bar{\Gamma} - \bar{X}$ and $\bar{\Gamma} - \bar{Y}$ along the [001]* and [110]* directions of the rutile primitive unit cell, respectively. Figure 1(c) presents the overview of the Fermi surface and band dispersions along the high symmetry $\bar{\Gamma} - \bar{X}$ and $\bar{\Gamma} - \bar{Y}$ lines measured using a photon energy value of 75 eV. For a better understanding, several other band cuts parallel to the $\bar{\Gamma} - \bar{X}$ and $\bar{\Gamma} - \bar{Y}$ are presented in Figure 1(e-j). The corresponding directions are labeled at the Fermi surface displayed in Figure 1(d). These cuts resolve the details of band dispersion over the whole BZ. The Fermi surface shows a circular feature centered at $\bar{\Gamma}$ ($k_x = k_y = 0$), surrounded by four other circular features (around $\bar{M}$). The same topography repeats over the adjacent BZ. The general Fermi surface features and band dispersions agree with recently reported ARPES data measured on $IrO_2$ thin films [30]. The high-quality single-crystal surface used for our measurement, however, significantly improves the data quality, and hence the resolution of the features. To further investigate the nature of electronic bands, we measured the Fermi maps using several different photon energies in the range 40 to 85 eV, which covers about one whole BZ periodicity along the $k_z$ direction. A strong $k_z$ dispersion within the measured photon energy range is observed as shown in Figure 2 (see also Supplementary Information). The evolution of Fermi surface distinguishes the bulk and surface features, e.g., the circular feature centered at Brillouin zone $\bar{M}$ point present across all photon energy values indicative of a surface state while the feature centered at $\bar{\Gamma}$ disperses with photon energy reminiscent of a bulk feature. On the other hand, the circular feature centered at the $\bar{\Gamma}$, which is most prominent at photon energy 75 eV, is clearly a bulk feature. Another arc like feature appears between the $\bar{\Gamma}$ and $\bar{X}$ points, and is most prominent at a photon energy of 85 eV, and therefore has bulk character. Likewise, a bulk circular feature around the $\bar{X}$ point appears only at 65 eV photon energy. A similar photon energy comparison of various band dispersions disentangles the surface and bulk features, which is further corroborated with surface and bulk calculations.

Recognizing spin-orbit derived features in experimental ARPES spectra is challenging, making theoretical support essential in the present case. To shed more light on our ARPES results, and to disentangle the role of SOC in generating the avoided band crossings, we have calculated the $IrO_2$ electronic structure using DFT and Green's functions calculations. The (110) band structures in general reveal an expected strong dependence on the perpendicular z component of the **k** vector, thus a direct comparison between theory and ARPES data is not straightforward. To better understand the simultaneous presence of surface and bulk features, we first examine the overall agreement for the measured and calculated surface states and less dispersive bulk features in Figure 3. We then qualitatively match the ARPES data taken at a specific photon energy with bulk calculations for a corresponding $k_z$ value (Figure 4). Figures 3(a) and 3(b) display the band structures along the short $\bar{M} - \bar{X}$ and $\bar{\Gamma} - \bar{Y}$ high-symmetry lines of the surface BZ, respectively (calculated by employing a semi-infinite surface model (left-hand panel) and measured at photon energy of 75 eV (right-hand panel)). The bands in Fig. 3(a) agree very well between experiment and theory, while the conical feature centered at $\bar{Y}$ (Figure 3(b)), given its bulk origin and dispersive character, only shows a qualitative agreement of the shape. The surface states of IrO$_2$(110) appear mainly along the $\bar{\Gamma} - \bar{X}$ and $\bar{Y} - \bar{M}$ directions. In Figures 3(c) and 3(d) we show the respective ARPES spectra (c.f. Figure 1(f) and 1(g) or 1(e)) with the most intense calculated surface states overlaid on top. This reveals a good agreement for the shape and position of the spectral features between experiment and theory. A comparison between theory and experiment of the Fermi surfaces (Figures 3(e) and 3(f)) is very satisfactory as well. The overall agreement between the experimental and calculated bands without considering correlation effects indicates that rutile IrO$_2$ shows modest correlation effects.

Figure 4(a) presents a DFT calculated bulk band structure along the $\bar{\Gamma} - \bar{X}$ line (at $k_z$ = 0), both with and without SOC (black and red lines). The electronic structure is calculated in a (110) supercell and contains states belonging to two different irreducible representations of the BZ, back-folded from the $\Gamma - Z$ and $A - M$ directions of the primitive cell. The ARPES spectrum measured within the surface BZ, shown in Figure 1, only reveals features corresponding to the $\Gamma - Z$ direction (plotted in Figure 4(a) as thick lines). To observe the bulk states lying along $A - M$ direction, we had to probe the second (110) surface Brillouin zone. We could not, however, identify clear bulk signatures due to the predominance of intense surface states. Furthermore, the non-symmorphicity of the rutile space group (*P4$_2$/mnm*), related to the existence of the screw axis, can further restrict the symmetries of the final states (transition matrix elements effect) that are observable, without destructive interference, in the angle-resolved photoemission process as discussed in Ref. [23].

**Discussion**

The feature of central interest is marked with the yellow ellipse; the avoided crossing at the Fermi level arises directly from the spin-orbit interaction and is extremely relevant for the spin Hall effect in IrO$_2$. As mentioned above and clearly explained in the work by Y. Sun *et al* [19], the large spin Hall conductivity in this material is associated to the existence of so-called Dirac nodal lines residing in $E(\mathbf{k})$ dispersion at (110) plane without SOC. Such lines form a continuous set of topologically non-trivial nodal points, which upon spin-orbit interaction, significantly contribute to the Berry curvature distribution in the **k**-space, and thus, to the large reported spin Hall conductivity. The principal nodal line in rutile IrO$_2$ merges the crossing points (without SOC) marked in Figure 4(a) as HP (hextuple point) and OP (octuple point) following the notation from Ref. [19]. In the ARPES experiment, however, the OP (lying along the $A - M$ direction of the primitive BZ) cannot be identified, due to above arguments above related to surface states; as such, the experimental observation of the gap opening along the whole nodal line in **k** space seems unfeasible. Rather, the avoided crossing feature associated with HP is explored. The expansion (Figure 4(b)) reveals its structure in closer detail; interestingly, apart from the spin-orbit split bands A and C, a double degenerate state B protruding through the gap is observed, which is a unique spectral feature and helps enable identification. All these bands have clear *Ir 5d* character hybridized with some O $2p_\pi$ character; following Ref. [21] for the Ir local frame, the main orbital character is given by $d_{xz}$ in case of band B and mixed $d_{yz}/d_{xz}$ at anti-crossing bands A/C, reminiscent of clean $d_{xz}$ ($d_{yz}$) of A(C) without SOC. This avoided crossing feature can be easily recognized in the calculated spectra of semi-infinite $IrO_2$(110), both in the bulk and surface projections (Figures 4(c) and 4(d), respectively). Moreover, despite numerous surface-like features dominating in all ARPES dispersions, we are clearly able to identify it. Figure 4(e) displays this avoided crossing feature measured at 60 eV photon energy, in excellent agreement with the bulk band structure calculated along $\bar{\Gamma} - \bar{X}$ direction ($\Gamma - Z$ in the primitive BZ) for $k_z$ = 0.1 (1/Å). In addition to the avoided-crossing, a pronounced surface state, certainly not reproduced in the superimposed bulk bands from DFT, is clearly visible in the surface *p*DOS (**k**, E) (Figure 4(d)), and further confirming the good agreement between experiment and theory.

In summary, we have experimentally investigated the electronic band structure and Fermi surface topography of $IrO_2$ by ARPES, using a high quality single crystal sample. By detailed photon energy dependent measurement, we unveiled a strong 3D character of the band structure as well as several clearly surface-derived features. With the support of first-principles calculations, we resolved various critical band structure features, including an avoided-crossing along the $\bar{\Gamma} - \bar{X}$ direction which, as revealed by our DFT calculations, results from spin-orbit coupling. Whilst the avoided crossing feature is itself extremely relevant for the reported spin Hall effect, its direct observation constitutes the first experimental signature of a recently suggested topological Dirac nodal line in the electronic structure of $IrO_2$ [19]. Our results open up exciting opportunities to further exploit this simple binary oxide for spintronic applications, and to understand the role that spin-physics plays in this material.

**Acknowledgements**
We would like to acknowledge and dedicate the manuscript to the late Professor Ying-Sheng Huang for providing the single crystals, which made this work possible. D.J.P. acknowledges support from the Royal Society (UF100105) and (UF150693). D.J.P. and A.R. acknowledge support from the EPSRC (EP/M013839/1). B.J.M. acknowledges support from the Royal Society (UF130329). The work at CNR-SPIN and CNR-IOM was performed within the framework of the nanoscience foundry and fine analysis (NFFA-MIUR Italy) project. J.S acknowledges the 'TO-BE' Cost Action for supporting the Short Term Scientific Mission. Part of the calculations has been performed in CINECA Supercomputing Center in Bologna. D. Di S. acknowledges the German Research Foundation (DFG-SFB 1170), the ERC-StG-336012-Thomale-TOPOLECTRICS.

**Captions for Figures**

**Figure 1**
**The electronic band structure and Fermi surfaces measured by ARPES. a-b.** Bulk and (110) surface Brillouin zones; in the former the yellow rectangle indicates the cleavage (110) plane. We define $\bar{\Gamma} - \bar{X}$ direction along [001]* and $\bar{\Gamma} - \bar{Y}$ direction along [110]*. **c.** 3D representation of the ARPES map including part of the Fermi surface and bands along $\bar{\Gamma} - \bar{X}$ and $\bar{\Gamma} - \bar{Y}$ directions **d.** the Fermi surface over a full Brillouin zone. Lines 1-6 correspond to particular band cuts which are displayed in panels **e-j**. These measurements were performed using photon energy ($h\nu$) = 75 eV at T = 78 K.

**Figure 2**
**Photon energy dependence of Fermi surface.** Fermi surfaces measured in the photon energy range from 50 eV to 85 eV with a step of 5 eV. The evolution of Fermi surface clearly distinguishes the bulk and surface features, e.g. the circular feature at M point present across all photon energy values indicative of a surface state and the feature centered at $\Gamma$ dispersing with photon energy and thus indicating a bulk character.

**Figure 3**
**ARPES and DFT Fermi surfaces measured and calculated on IrO$_2$ (110) a.** Left: Density of states projected on surface layers along the $\bar{M} - \bar{X} - \bar{M}$ line obtained via DFT and Green's functions calculations within semi-infinite surface model. Right: Same map obtained via ARPES at photon energy 75 eV. The spectra are not superimposed since the main conical feature at $\bar{X}$ has bulk-like character, thus depends on $k_z$ and on the specific choice of photon energy. **b.** Same as **a** for along $\bar{\Gamma} - \bar{Y} - \bar{\Gamma}$ direction. **c.** Electronic band structure along $\bar{\Gamma} - \bar{X} - \bar{\Gamma}$ obtained via ARPES at photon energy 75 eV. Only the left-hand side corresponds to the directly measured spectrum, which has been further replicated and reflected in order to obtain a symmetric image. Theoretical data are superimposed in blue on top of the experimental ARPES spectra. The contrast of the former is adjusted to show only the most intense (surface) features. The calculated map is shifted by 50 meV towards lower binding energy in order to achieve better agreement with experiment. **d.** Same as **c** for along $\bar{Y} - \bar{M} - \bar{Y}$ direction. **e.** Fermi surface measured via ARPES at photon energy value of 75 eV. The part inside the yellow rectangle corresponds to data measured directly with ARPES, while the rest of the image consists of the correspondingly reflected replicas. The symmetrization aims at better comparison between experimental and theoretical $\Gamma$- centered Fermi surface. **f.** Same as **e** with the overlaid theoretical results calculated via Green's functions approach for the bulk projection of semi-infinite IrO2(110). The theoretical Fermi surfaces have been calculated at the energy $E$ = - 50 meV in agreement with panels **c** and **d**.

**Figure 4**.
**Band structure dispersion measured and calculated for the $\bar{\Gamma} - \bar{X} - \bar{\Gamma}$ direction a**. Bulk band structure along $\bar{\Gamma} - \bar{X} - \bar{\Gamma}$ calculated via DFT in (110) supercell at $k_z$ = 0. The states plotted with thick lines correspond to $\Gamma - Z$ direction of the primitive unit cell, while the bands marked as thinner are back folded from $A - M$ direction and are unlikely to be observed in ARPES experiment (see text). Black and red lines correspond to the calculations performed without and with SOC, respectively. Yellow ellipse marks a region with SOC-induced avoided crossing (band gap opening). **b.** Zoom in of the features marked with yellow ellipse in panel **a**, with and without SOC. In the latter three doubly degenerate bands, A, B, C cross at one hextuple point (HP). Bands A and B are identical (two lines on top of each other). When SOC is switched on, a large gap opens and bands A and C change their character, while band B remains unaffected. **c.** Electronic band structure corresponding to bulk projection of IrO$_2$(110) semi-infinite system along $\bar{\Gamma} - \bar{X} - \bar{\Gamma}$ calculated via Green's functions approach. Red lines superimposed on the left-hand side are identical to those at panel **a**. **d.** Same as **c.** for surface projection. The green arrow indicates one of the intense surface states. **e.** Electronic band structure along $\bar{\Gamma} - \bar{X} - \bar{\Gamma}$ measured with ARPES at photon energy of 60 eV symmetrized with respect to $\Gamma$. The red lines represent bands calculated at $k_z \approx$ 0.1 (1/Å) roughly corresponding to the selected photon energy in the ARPES map. Again, we present only the bands back folded from $\Gamma - Z$ direction. The surface state marked by an arrow corresponds to the one in panel **d**.

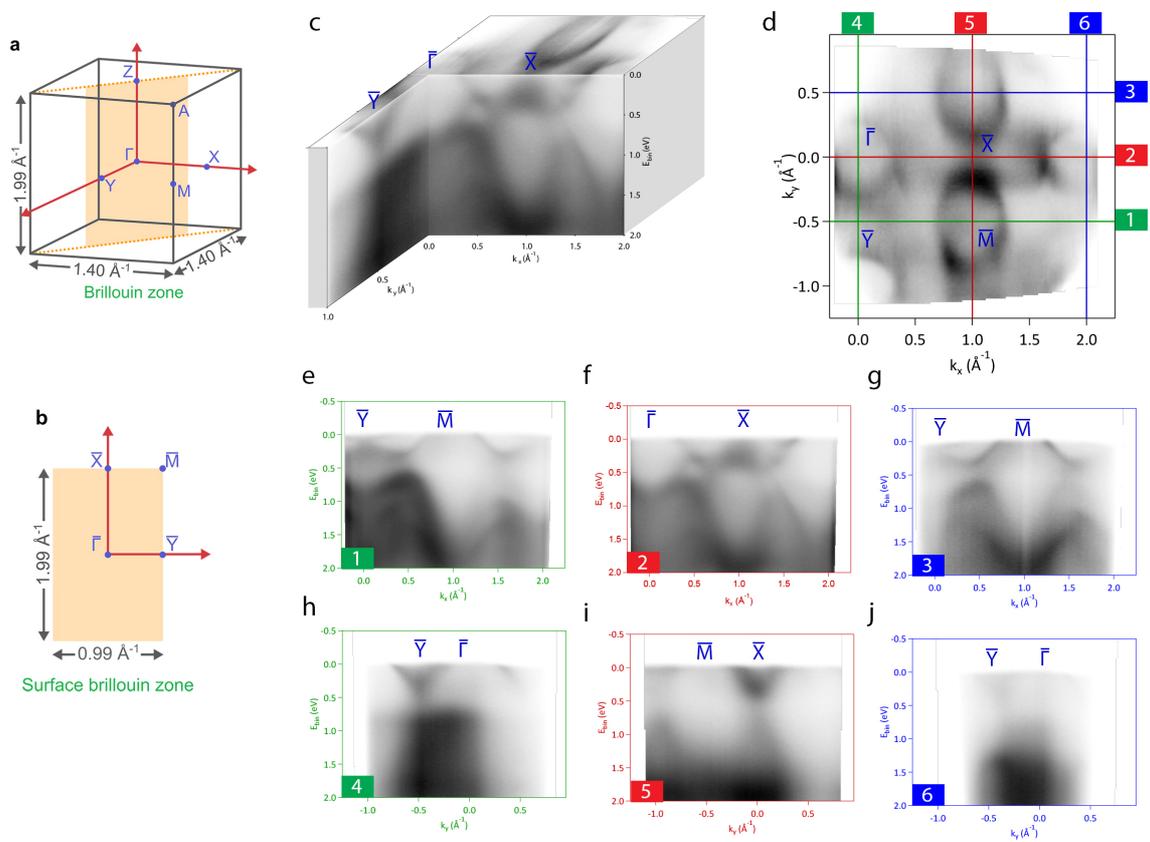

**Figure 1**

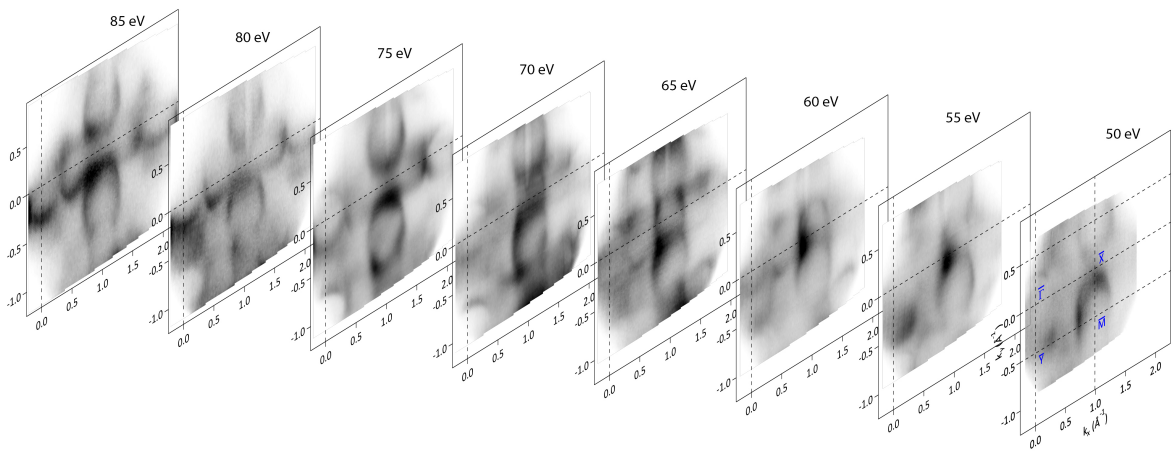

**Figure 2**

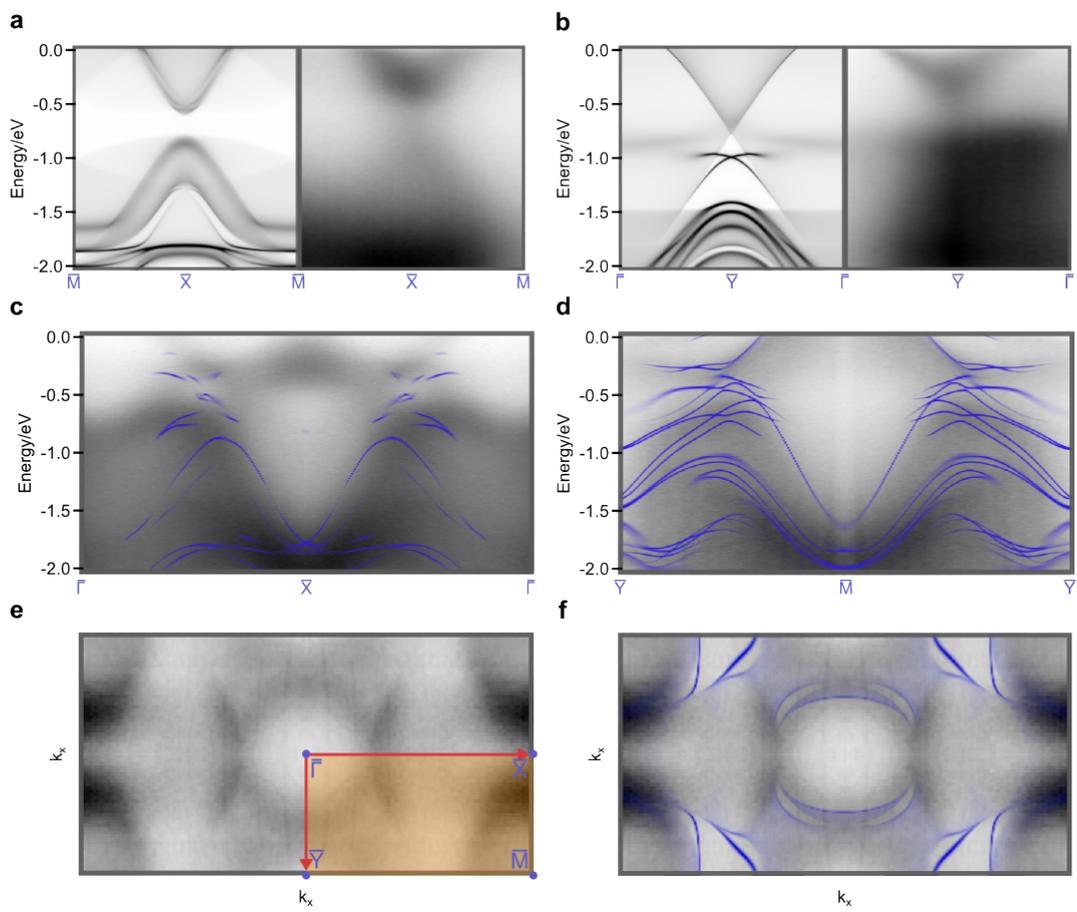

**Figure 3**

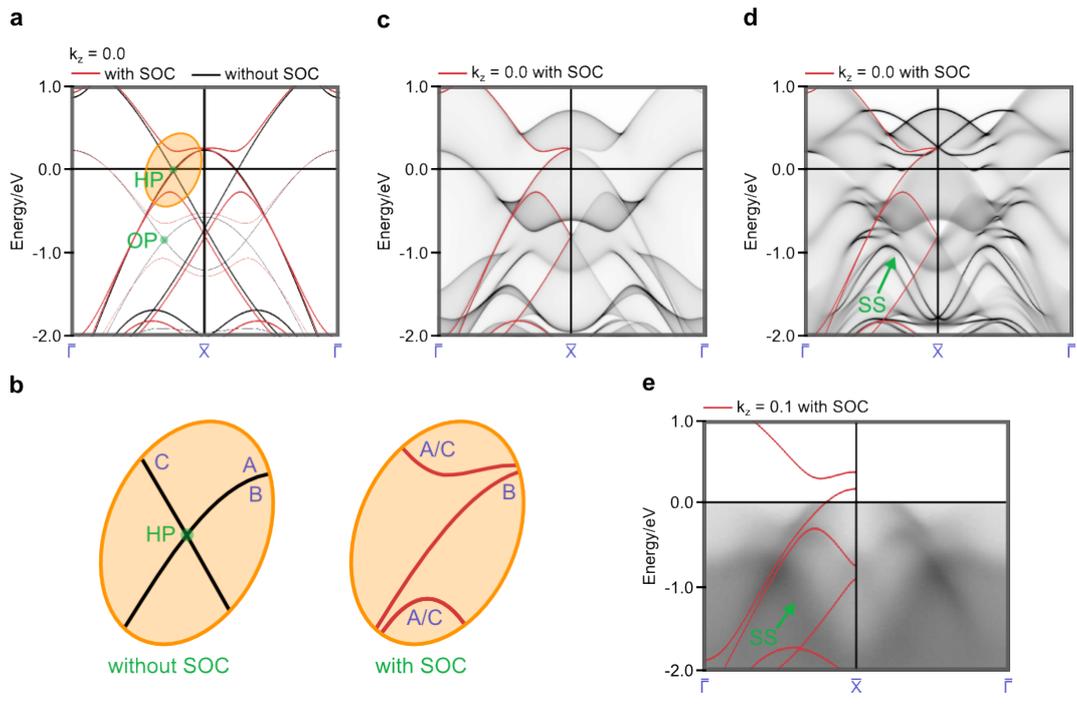

**Figure 4**